\documentclass[aip,
 jmp, bmf,
 sd,
 rsi,
 amsmath,amssymb,
preprint,
 reprint,
author-year,
author-numerical,
Conference Proceedings
]{revtex4-1}
\usepackage{epsfig}
\usepackage[normalem]{ulem}
\usepackage{graphics}
\usepackage{lipsum}
\usepackage{amsmath}
\usepackage{sidecap}
\usepackage{mathtools}
\usepackage{graphicx} 
\usepackage{dcolumn} 
\usepackage{bm}
\usepackage{xcolor,colortbl}
\usepackage{color, colortbl}
\definecolor{darkgray}{rgb}{0.66, 0.66, 0.66}
\definecolor{yellow-green}{rgb}{0.6, 0.8, 0.2}
\definecolor{deeppink}{rgb}{1.0, 0.08, 0.58}
\definecolor{darkviolet}{rgb}{0.58, 0.0, 0.83}

\definecolor{darkcyan}{rgb}{0.0, 0.55, 0.55}
\usepackage{sidecap}  
\begin{document}
\title{Frequency enhancement and power tunability in tilted polarizer spin-torque nano oscillator}
\author{R. Arun$^{1}$}
\email{arunbdu@gmail.com}
\author{R. Gopal$^{2}$}
\email{gopalphysics@gmail.com}
\author{V.~K.~Chandrasekar$^{2}$}
\email{chandru25nld@gmail.com}
\author{M.~Lakshmanan$^1$}
\email{lakshman.cnld@gmail.com}
 
\affiliation
{
$^{1}$Department of Nonlinear Dynamics, School of Physics, Bharathidasan University, Tiruchirapalli-620024, India\\
$^{2}$Centre for Nonlinear Science \& Engineering, School of Electrical \& Electronics Engineering, SASTRA Deemed University, Thanjavur- 613 401, India. \\
}

\begin{abstract}
In the absence of an applied magnetic field, a spin-torque nano oscillator(STNO) with a tilted polarizer is studied using numerical simulation of the associated Landau-Lifshitz-Gilbert-Slonczewski equation. We find considerable enhancement of frequency by tilting the polarizer out-of-plane appropriately.  Also, we observe improved tunability of frequency of oscillations from $\sim$15 GHz to $\sim$75 GHz and increment in the power spectral density  by current and tilt angle.  In addition, our findings and insights pave a simple way for nanoscale level microwave generators to be implemented.

\end{abstract}


\maketitle
\section{Introduction}
Spin-torque nano oscillator(STNO) is a nanosized device that consists of two ferromagnetic layers, one with fixed magnetization and the other with variable magnetization direction, separated by a thin non-magnetic conductive layer~\cite{kiselev:00,rippard:04,krivorotov:05,slavin:06}.  When a current is passed through the STNO the electrons get spin polarized and excert a spin-transfer torque on the magnetization of the free layer~\cite{slon:96,berger:96,slon:02}. While the current density is above a threshold value, the spin transfer torque oscillates the magnetization of the free layer continuously and these oscillations are transformed into microwave voltage oscillations by magnetoresistive effects~\cite{sun:00,ralph:08}.  Also, these STNOs are capable of generating microwave signals with a broad range of frequencies in gigahertz scale~\cite{ralph:08,pufall:05,silva:08}. STNOs have many attractive features such as ultra-small dimension~\cite{slon:96}, wide tunability range~\cite{bonetti:09}, compatibility during the semiconductor fabrication process and very high modulation rates~\cite{mudali:10,mudali:11,mudali:11:1}. Another important feature of STNOs is the ability to tune their oscillation frequencies with power enhancement, which is essential for wireless communication devices~\cite{Tare:19}. STNOs have shown tremendous growth towards experimental realization with large operating frequency~\cite{mohseni:11}.   Nevertheless, STNO devices face issues such as requirment of large static magnetic field and  generation of microwave with low frequency and output power.  

A lot of techniques have been implemented by various researchers to achieve the required power, which include utilizing magnetic tunnel junction in single STNO~\cite{deac:08,nazarov:08,pala} and synchronizing arrays of STNOs to produce enhanced coherent microwave oscillations~\cite{grollier:06,persson:07,li:11,li:10,li:11:1,subash1,subash2,turtle,gopal,jensen}. In addition to the above, it has been observed that an STNO with perpendicularly polarized pinned layer may drive an in-plane magnetization of free layer into an out-of-plane precessional state even in the absence of applied magnetic field~\cite{redon:03,lee:05}. Additionally, the most widely adopted zero field operations also include STNOs with perpendicular magnetic anisotrophy in free and pinned layers~\cite{rippard:10,mohseni:11}, dual spin polarizers~\cite{lee:11}, vortex oscillators~\cite{pribiag:07,pribiag:09,finocchio:10,locatelli:11,dussaux:11}, current induced Oersted field in STNO~\cite{bhoomeeswaran:18}, wavy torque spin-torque oscillators~\cite{boulle:07}, field-like torque~\cite{tani:14},  achievement of doubling of output signal frequency and power with the help of  dual free STNOs~\cite{pro:13}, study of synthetic-ferro magnetic spin-torque oscillator in the absence of external field~\cite{zhou:13}, and tilted polarizer STNOs~\cite{bhoomeeswaran:18,zhou1,zhou4,zhou5,he:09,he:10,wang:11,zha:09,zha:09:01,zha:09:02,zhou:12,gang:15,cui}. 

A tilted polarizer STNO(TP-STNO) is a STNO where the magnetization of its pinned layer can be tilted out of the film plane by an angle designated as tilt angle.  When the pinned layer is tilted, it exhibits two components of magnetization, in-plane and out-of-plane. The spin current polarized by the out-of-plane component of the magnetization drives the magnetization of the free layer into steady state precessional motion in the absence of magnetic field. 

 Initially, the TP-STNO has been studied by Zhou $et~al.$ and the maximum operating frequency of TP-STNO up to 29 GHz as a function of drive current has been reported for Py material~\cite{zhou1}. Further, the specific magnetization dynamics of TP-STNO as a function of the device's drive current and fixed layer angle as well as its related microwave frequency and efficient magnetoresistance have been studied~\cite{zhou4,zhou5}.  In continuation of the above studies, an intense interest has also been shown on studying such TP-STNO based devices with the aid of design and experimental fabrication process~\cite{zha:09,zha:09:01,zha:09:02}. In particular, Zhou and Gang have reported a full-scale micromagnetic study on the dynamics of TP-STNO~\cite{zhou:12,gang:15}.  Based on the analytical theory and macrospin simulations Zhou $et~al.$ have reported the hysteretic switching between bistable states in TP-STNO~\cite{zhou:12}.  Gang $et~al.$ have shown the dependence of precession frequency on material parameters of TP-STNO and acheived the frequencies ranging from 1.8 GHz to 41.2 GHz~\cite{gang:15}.  The range of drive current density for which the TP-STNO exhibits stable oscillations in the presence of thermal noise for different tilt angles has been identified and reported by Cui $et~al.$~\cite{cui}.  The above mentioned studies show the significant advantage of TP-STNO, functioning without external magnetic field, in order to maintain or generate microwave signal with high frequency and power. In the earlier studies on TP-STNO, the tunability of high frequency in the zero-field operation has been addressed by the spin torque amplitude with in-plane asymmetric angular dependence~\cite{zhou1,zhou4,zhou5}.  Further enhancement of frequency(above 50 GHz) and power in TP-STNO along with a systematic study on both static and dynamical aspects of TP-STNO is required to be addressed for the development of nanoscale structured current driven TP-STNO devices.

\begin{figure*}[htp]
 	\centering\includegraphics[angle=0,width=1.0\linewidth]{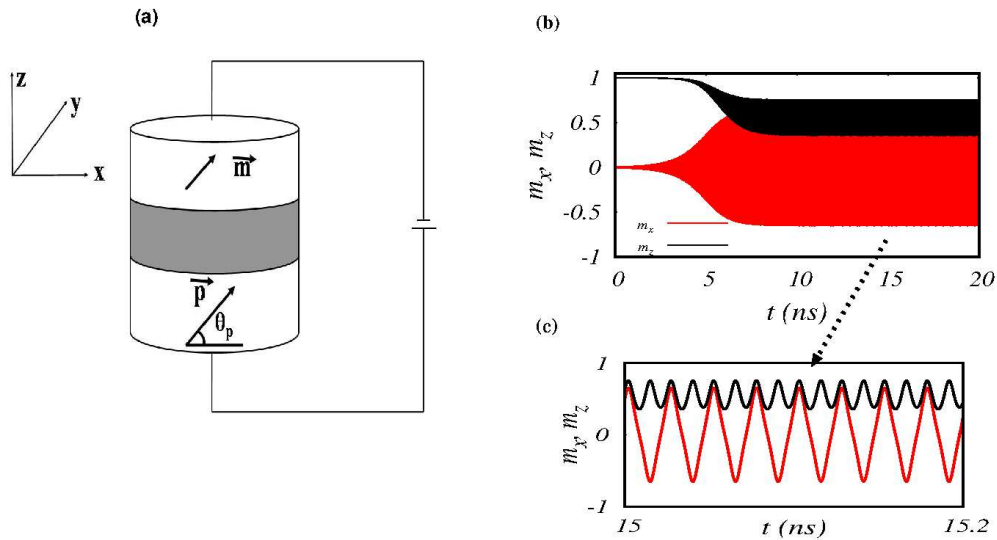}
 	\caption{(Color online) (a)Illustration of the typical structure of TP-STNO and (b) Time evolution of $m_x$ and $m_z$ when $I$ = -2.5 mA and $\theta_p$ = 45$^\circ$.  (c) Magnification of the evolution of $m_x$ and $m_y$ in a short range of time $t \in$[15.0 ns, 15.2 ns]}
\label{fig1}
\end{figure*}

Therefore, in our present study, we report the high frequency magnetization oscillations with large tunability and power enhancement in a TP-STNO having inplane magnetized free layer  by considering the spin torque amplitude with both in-plane and out-of-plane asymmetric angular dependence. We have derived the relation between the  current and the magnetization precession frequency of the TP-STNO under zero applied field. The validity of the numerical simulation is closely confirmed with the analytical solution. The paper is organized as follows. Section II addresses a detailed geometry of the TP-STNO and its governing equation, namely the Landau-Lifshitz-Gilbert-Slonczewski(LLGS) equation, for the dynamics of magnetization precession. The dynamics of the free layer magnetization for frequency tunabilty by varying tilt angles through numerical and analytical studies are presented in Sec III. Finally, concluding remarks are made in Section IV.   In the Appendices, we obtain the analytic form of steady states of magnetization for different values of spin current, impact of damping, saturation magnetization and thermal noise on the frequency of oscillations.

\begin{figure}
  	\centering\includegraphics[angle=0,width=0.7\linewidth]{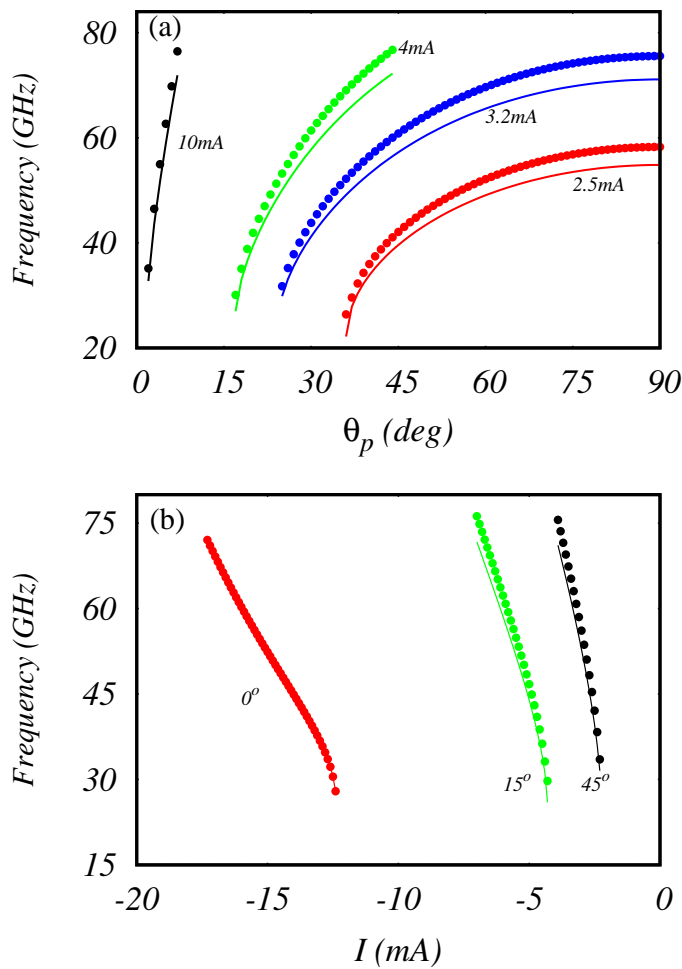}
  	\caption{(Color online) (a) Tunability of frequency at different currents with respect to tilt angle and (b) Tunability of frequency at different tilt angles with respect to current.}
  	\label{fig2}
\end{figure}

    \begin{figure*}
    	\centering\includegraphics[angle=0,width=1\linewidth]{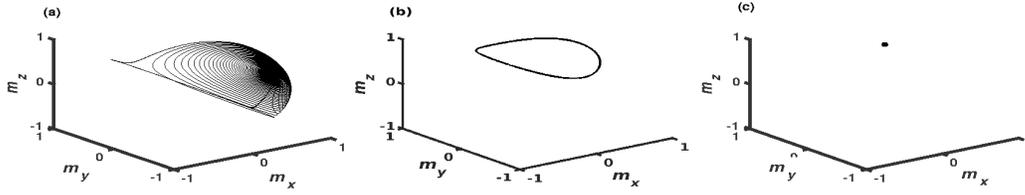}
    	\caption{(Color online) Trajectory of magnetization motion at (a) -2.2 mA, (b) -3.0 mA and (c) -5.0 mA when $\theta_{p}=45^{0}$.}
    	\label{fig3}
    \end{figure*}

\section{Model}
 
We consider a model structure of TP-STNO which is shown in Fig.\ref{fig1}(a). It consists of a free layer, a fixed layer and a spacer layer between them. The free and fixed layers are the ferromagnetic layers, where the former is allowed to change the direction of its magnetization and the later is not.  Spacer layer is a nonmagnetic but conducting layer. The unit vector of the free layer's magnetization is given by ${\bf m} = m_{x}{\bf e}_{x}+m_{y}{\bf e}_{y}+m_{z}{\bf e}_{z}$.  ${\bf e}_{x}$,${\bf e}_{y}$ and ${\bf e}_{z}$ are considerd as the unit vectors along positive $x$,$y$ and $z$ directions respectively. The plane of the free layer is taken perpendicular to  ${\bf e}_z$.  The LLGS equation that governs the dynamics of the free layer's magnetization is given by
\begin{align}
\frac{d{\bf m}}{dt}=-\gamma ~{\bf m}\times{\bf H}_{eff}+ \alpha ~{\bf m}\times\frac{d{\bf m}}{dt}+\gamma ~H_{S} ~{\bf m}\times ({\bf m}\times{\bf p}), \label{llgs}
\end{align}
where $\gamma$ is the gyromagnetic ratio, $\alpha$ is the Gilbert damping parameter, ${\bf H}_{eff}$ is the effective field and the term $H_{S}$ is the strength of the Slonczewski spin-transfer torque~\cite{slon:02} with magnitude
\begin{equation}
H_{S} = \frac{\hbar\eta I}{2eM_s V(1+\lambda{\bf m}.{\bf p})}.\label{Hs}
\end{equation}
 In Eq.\eqref{Hs}  $\hbar = h/2\pi$ is the reduced Planck's constant ($h$ - Planck's constant), $V$ is the volume of the free layer, $M_s$ is the saturation magnetization, $e$ is the electron charge,  $I$ is the total current passing through the free layer,  $\eta$ and $\lambda$ are dimensionless parameters which determine the magnitude of the spin polarization and the angular dependence of the spin transfer torque, respectively.

 The effective field ${\bf H}_{eff}$ can be expressed as ${\bf H}_{eff}=H_k m_x {\bf e}_x - 4\pi M_s m_{z}{\bf e}_z$ which includes magnetocrystalline anisotropy field $H_k$ and demagnetization field $4\pi M_s$. Further, the magnetization of the pinned layer is considered to be tilted in the $xz$-plane by the tilt angle $\theta_p$ from positive $x$-axis and the unit vector of the pinned layer's magnetization is given by ${\bf p} =~ \cos\theta_p  ~{\bf e}_x+ \sin\theta_p ~{\bf e}_z$.The current in Eq.\eqref{Hs} is considered as positive or negative when it flows across the STNO from pinned layer to free layer or free layer to pinned layer respectively.  The material parameters $M_s = 1448.3$ emu/c.c., $H_k = 18.6$ kOe, $\eta$ = 0.54, $\lambda$ = $\eta^2$, $\gamma$ = 17.64 Mrad/(Oe s), $\alpha$ = 0.005, and $V=\pi \times 60 \times 60 \times 2$ nm$^3$ for cobalt material have been adopted from Ref.\cite{tani:18}.

 \section{Results and Discussion}

 The oscillations of ${\bf m}$ around the z-axis due to tilted polarizer is confirmed in Fig.\ref{fig1}(b) by plotting $m_x$ and $m_z$ with respect to time for $I$ = -2.5 mA and $\theta_p$ = 45$^\circ$.  The inset figure in Fig.\ref{fig1}(b) confirms the oscillations of $m_x$ and $m_z$ with respect to time.  It has been numerically verified that the oscillations are not exhibited when the current is applied along the positive direction. The frequency of oscillations in terms of $I$ and $\theta_p$ is determined from Eq.\eqref{llgs} after transforming it into spherical polar co-ordinates
 ${\bf m} = (\sin\theta \cos\phi, \sin\theta \sin\phi, \cos\theta)$ as follows:
 \begin{align}
 &\frac{(1+\alpha^2)}{\gamma}\frac{d\theta}{dt} ~=~ H_k \sin\theta \cos\phi (\alpha \cos\phi \cos\theta-\sin\phi) \nonumber\\ &+ 4\pi M_s \alpha \sin\theta \cos\theta- H_S \cos\theta_p ~[\cos\phi\cos\theta+\alpha\sin\phi] \nonumber\\ &+  H_S \sin\theta_p ~\sin\theta,\label{polar_theta}\\
 &\frac{(1+\alpha^2)}{\gamma}\sin\theta\frac{d\phi}{dt} ~=~ -H_k \sin\theta\cos\phi(\alpha\sin\phi+\cos\theta\cos\phi)\nonumber\\ &-4\pi M_s \cos\theta\sin\theta+H_S \cos\theta_p ~[\sin\phi-\alpha\cos\theta\cos\phi] \nonumber\\ &+ H_S \sin\theta_p ~\alpha \sin\theta. \label{polar_phi}
 \end{align}
 Since the amplitude of $m_z$ is small(see Fig.\ref{fig1}(b)), $\theta$ can be approximated as constant and $d\phi/dt = 2\pi f = 2\pi/T$, where $f$ is the frequency and $T$ is the time period  of the oscillations. By integrating Eq.\eqref{polar_phi} with respect to time from 0 to $nT$, where $n$ is the number of oscillations, we can derive the frequency as,
 \begin{align}
 &f=~ \frac{\gamma}{2\pi(1+\alpha^2)}\left\{-\left(\frac{H_k}{2}+4\pi M_s\right)\cos\theta\right.\nonumber\\&\left.+
 \alpha H_{S}\left[Q~{\sin\theta_p}-\frac{\cos\theta }{\lambda\sin^2\theta}\left(1-Q~({1+\lambda \sin\theta_p \cos\theta}) \right) \right]\right\},\label{anafreq}
 \end{align}
 where $Q=[{1+2\lambda\sin\theta_p\cos\theta+\lambda^2(\sin^2\theta_p-\sin^2\theta)}]^{-1/2}$.

Equation \eqref{anafreq} qualitatively denotes the dependence of frequency $f$ with respect to tilt angle $\theta_{p}$ and flow of current $I$.   Figs.\ref{fig2}(a) and (b) show the variation of frequency $f$ versus $\theta_{p}$ and $I$ respectively. In Figs.\ref{fig2}, the solid line corresponds to numerically computed frequency and bullets correspond to analytically computed frequency using Eq.\eqref{anafreq}. From Figs.\ref{fig2}(a) and (b) it is clearly observed that the frequency of the magnetization precession enhances by increasing the values of $\theta_{p}$ and $I$ respectively. Here, the initial conditions for ${\bf m}$ are chosen to be very close to ${\bf e}_z$.   In addition to this choise of initial condition, we find that oscillations are also possible when the initial state of the magnetization is taken upto 45$^\circ$ from positive z-direction in the xz-plane as well as near negative y-direction. For other choises of initial conditions the magnetization reaches steady state along the positive x-direction without showing any stable oscillations. Since the frequencies corresponding to $\theta_p$ from 90$^\circ$ to 180$^\circ$ are symmetrical with the frequencies corresponding to $\theta_p$ from 0$^\circ$ to 90$^\circ$,  they are not shown in Fig.\ref{fig2}(a). Further, Figs.\ref{fig2}(a) and (b) show that the current required for the maximum frequency can be reduced when the pinned layer's magnetization is tilted out of the plane.  The small discrepancy between the analytical(bullets) and numerical(lines) frequencies in Figs.\ref{fig2} arises due to the small amplitude approximation of $m_z$. From Figs.\ref{fig2} we can clearly observe that the frequency can be tuned from $\sim$25 GHz to $\sim$75 GHz by varying the tilt angle and the magnitude of current.  

In order to depict the detailed magnetic precession dynamics of the TP-STNO, we carried out macromagnetic simulations for different values of current which is shown in Fig.\ref{fig3}. Figures \ref{fig3}(a), (b) and (c) confirm that the magnetization exhibits steady state motion towards ${\bf e}_x$, steady precession  about $z$-axis and steady state towards ${\bf e}_z$ at different currents -2.2 mA, -3.0 mA and -5.0 mA respectively. Here, one may note that the magnetization switches between oscillatory and steady state depending upon the value of current. Also it indicates that the magnetization precession is possible only when the magnitude of the current is between two critical values, denoted as $I^c_{max}$ and $I^c_{min}$.  Beyond this region of current the magnetization exhibits steady state. When the magnitude of current is below $I^c_{min}$, the magnetization flips to the $xy$-plane and reaches steady state near ${\bf e}_x$(or $-{\bf e}_x$).  Similarly, when the magnitude of current is above $I^c_{max}$ the magnetization approaches the steady state close to ${\bf e}_z$. For instance, the values of $I^c_{max}$ and $I^c_{min}$ are identified numerically as -3.9 mA and -2.3 mA respectively when $\theta_p$ is 45$^\circ$. 

To obtain a deeper insight into the  stable precessions and steady state, the current required for the  oscillation corresponding to $\theta$ can be estimated by incorporating the condition $\frac{1}{nT}\int_0^{nT} \frac{d\theta}{dt}~dt=0$ on Eq.\eqref{polar_theta} as,
 \begin{align}
 I = \frac{(2 e M_s V/\hbar\eta) \alpha \cos\theta (\frac{H_k}{2}+4\pi M_s)}{\frac{\cos\theta}{\lambda\sin^2\theta}[1-(1+\lambda \sin\theta_p \cos\theta)R]-\sin\theta_p~R},\label{criticalcurrent}
 \end{align}
 where,
 $R = [(1+\lambda \sin\theta_p \cos\theta)^2- \lambda^2\cos^2\theta_p \sin^2\theta ]^{-1/2}$.  From Eq.\eqref{criticalcurrent} we can derive the $I_{max}^c$ by taking $\theta \rightarrow 0$ as
 \begin{align}
 I_{max}^c = \frac{-2 e M_s V \alpha (1+\lambda \sin\theta_p)(\frac{H_k}{2}+4\pi M_s)}{\hbar \eta \left[\sin\theta_p+\frac{\lambda \cos^2\theta_p}{2(1+\lambda \sin\theta_p)}\right]}.\label{Icmax}
 \end{align}
 Eq.\eqref{Icmax} gives the magnitude of current above which TP-STNO exhibits steady state only.

\begin{figure}
\centering\includegraphics[angle=0,width=1.0\linewidth]{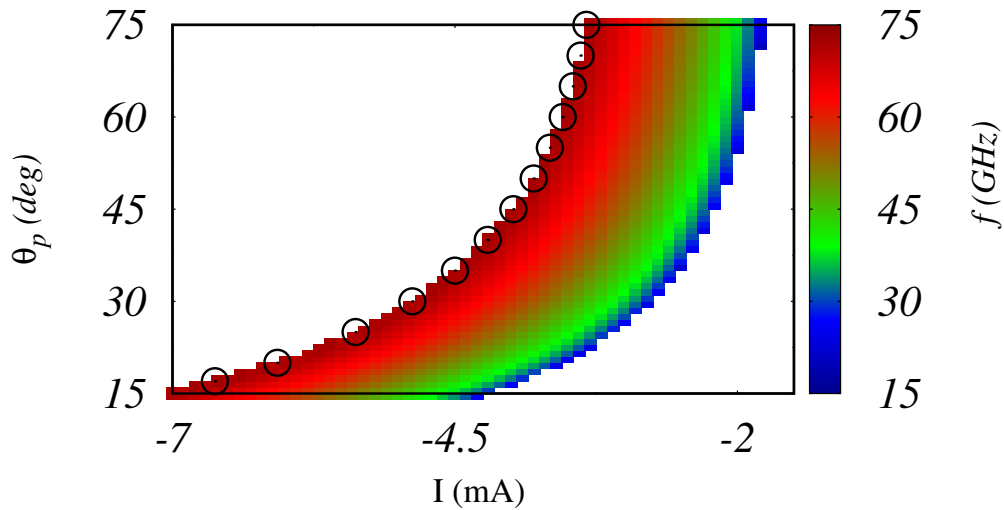}
\caption{(Color online) The region of current and tilt angle for which the oscillations are possible.  The white regions are corresponding to steady state. The black circles are the values of $I_{max}^c$ plotted from Eq.\eqref{Icmax}.}
\label{fig4}
\end{figure}

The possible values of $\theta_{p}$ and $I$ corresponding to the steady state and oscillatory motions of magnetization are depicted by plotting the frequency in Fig.\ref{fig4} with respect to $\theta_{p}$ and $I$. Fig.\ref{fig4} confirms that the magnitude of current required for a given frequency reduces when the tilt angle is increased.   Also, we can understand that the frequency can be enhanced when the tilt angle(current) is increased while the current(tilt angle) is fixed.  And the frequency of oscillations is enhanced up to $\sim$50 GHz by tuning the tilt angle and current. The black circles in Fig.\ref{fig4} represent the values of $I_{max}^c$ obtained from Eq.\eqref{Icmax}. 
 
Since the magnetization settles into the xy-plane and aligns closely with positive or negative $x$ direction when the current is below $I_{min}^c$, the steady state values can be identified from Eqs.\eqref{polar_theta} and \eqref{polar_phi} as (see Appendices A and B)
\begin{align}
\theta^* = & \frac{\pi}{2}\label{ss1}\\
\phi^* = \frac{H_{S} \sin\theta_p}{H_k (1+\lambda\cos\theta_p)}~~&(\textrm{or})~~\pi+\frac{H_{S} \sin\theta_p}{H_k (1-\lambda\cos\theta_p)}, \label{ss2}
\end{align}
where, $H_{S}=\hbar \eta I/2eM_s V$. Similarly, the magnetization aligns closely with positive $z$ direction and settles in the fourth quadrant of $xy$-plane when the magnitude of current is above $I_{max}^c$ and the corresponding steady state points are also obtained from Eqs.\eqref{polar_theta} and \eqref{polar_phi} as
\begin{align}
\theta^* &= \frac{-H_{S}\cos\theta_p}{4\pi M_s(1+\lambda\sin\theta_p)},\label{ss3}\\
\phi^*  &= \frac{3\pi}{2}-\frac{H_{S} \sin\theta_p }{(H_k+4\pi M_s) (1+\lambda\sin\theta_p)}.\label{ss4}
\end{align}
Eqs.\eqref{ss1} and \eqref{ss2} represent the steady state points when the magnitude of current is below $I_c^{min}$ and the Eqs. \eqref{ss3} and \eqref{ss4} represent the steady state points when the magnitude of current is above $I_{max}^c$.

\begin{figure}[htp]
	\centering\includegraphics[angle=0,width=1.0\linewidth]{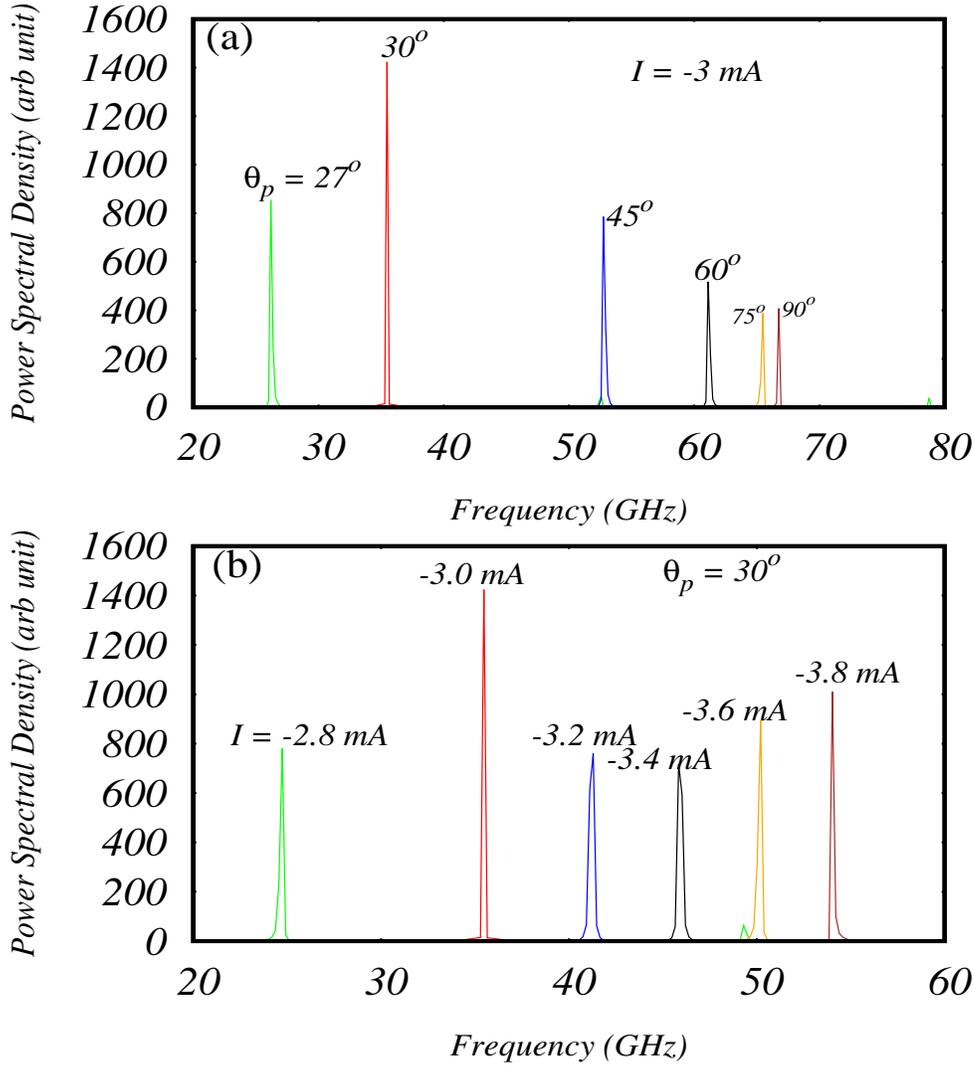}
	\caption{(Color online) The power spectral densities at different (a) $\theta_p$ when $I$ = -3 mA and (b) $I$ when $\theta_p$ = 30$^\circ$ }
	\label{fig5}
\end{figure}

Finally, the power spectral densities corresponding to the current $-3.0$ mA and tilt angle 30$^\circ$ are plotted in Figs.\ref{fig5}(a) and (b) respectively. Fig.\ref{fig5}(a) shows that the peak height increases when the magnetization of the pinned layer is tilted out-of-the plane, which confirms that the power can be enhanced by tilted polarizer.  The enhancement of peak height at large frequencies in Fig.\ref{fig5}(b) when the magnitude of current is increased implies that the power can be enhanced at small as well as large frequencies by tilting the magnetization of the pinned layer.

 Also in Appendix C we point out the impact of the damping parameter $\alpha$  and saturation magnetization $M_s$ for cobalt material.  Further in Appendix D we point out that thermalization does not lead to any appreciable change in the frequency of oscillations.

\section{Conclusions}
In summary,  spin-transfer torque induced magnetization dynamics in a TP-STNO consisting of an in-plane magnetized free layer and a pinned layer  with the tunability of its magnetization direction has been analytically and numerically studied by solving the governing LLGS equation in the absence of external magnetic field.  We find that the frequency of oscillations can be tuned from $\sim$15 GHz to $\sim$75 GHz by tilting the magnetization of the pinned layer into out-of-plane.  Also, we have found that the current required for the maximum frequency can be reduced when the pinned layer's magnetization is tilted.  Further, we have observed the switching of magnetization between precessional state and steady state takes place depending upon the value of current. Therefore, the present study of the TP-STNO yields the combined advantage of zero field operation and high frequency signal which can give potential applications for future wireless communications.

\vspace{-0.5cm}
\section*{Acknowledgements}
The work of V.K.C. forms part of a research project sponsored by CSIR Project No. 03/ 1444/18/EMR II. M.L. wishes to thank the Department of Science and Technology for the award of a SERB Distinguished Fellowship under Grant No.SB/DF/04/2017 in which R. Arun is supported by a Research Associateship.

\section*{Appendix}
\subsection{Steady states when $I < I_{min}^c$}
When the current is below $I_{min}^c$, the magnetization flips to the $xy$-plane and may settle near positive or negative $x$ direction depending upon the initial conditions. The steady state points are identified as $\theta^*=\pi/2$ and $\phi^*=\delta\phi ~\rm{or}~\pi+\delta\phi$, where $\delta\phi\approx 0$. The quantity $\delta\phi$ can be derived by substituting $\frac{d\phi}{dt}=\frac{d\theta}{dt}=0$, $\cos\phi=\pm1$ and $\sin\phi=\pm \delta\phi$ in Eqs.\eqref{polar_theta} and \eqref{polar_phi} as
\begin{align}
-H_k \delta\phi \mp \frac{H_{S}\alpha\cos\theta_p \delta\phi}{1\pm\lambda\cos\theta_p}+\frac{H_{S}\sin\theta_p}{1\pm\lambda\cos\theta_p}&=0\label{a1},\\
-\alpha H_k \delta\phi \pm \frac{H_{S}\cos\theta_p \delta\phi}{1\pm\lambda\cos\theta_p}+\frac{H_{S}\alpha\sin\theta_p}{1\pm\lambda\cos\theta_p}&=0\label{a2}.
\end{align}
From Eqs.\eqref{a1} and \eqref{a2} we can derive,
\begin{align}
\delta\phi = \frac{H_{S} \sin\theta_p}{H_k(1\pm \lambda\cos\theta_p)}\label{a3}.
\end{align}
Hence,
\begin{align}
\phi^* = \frac{H_{S} \sin\theta_p}{H_k(1+ \lambda\cos\theta_p)} ~~\rm{(or)}~~\pi+\frac{H_{S} \sin\theta_p}{H_k(1- \lambda\cos\theta_p)}\label{a4}
\end{align}
\subsection{Steady states when $I > I_{max}^c$}
Similarly, the steady states are identified as $\theta^*\approx 0$ and $\phi^*=\frac{3\pi}{2}+\delta\phi'$ when the current is above $I_{max}^c$. $\theta^*$ can be derived by substituting $\frac{d\phi}{dt}=\frac{d\theta}{dt}=0$, $\cos\phi=\delta\phi'$ and $\sin\phi=-1$ in Eqs.\eqref{polar_theta} and \eqref{polar_phi} and after omitting terms with ${\delta\phi}^2$ as
\begin{align}
[(H_k \delta\phi' + 4\pi M_s  \alpha)P+H_{S}\sin\theta_p]\theta^*=H_{S}(\delta\phi'-\alpha)\cos\theta_p,\label{a5}\\
[(\alpha H_k \delta\phi'-4\pi M_s)P+H_{S}\alpha\sin\theta_p]\theta^*=H_{S}(1+\alpha\delta\phi')\cos\theta_p,\label{a6}
\end{align}
where, $P=1+\lambda \theta^* \cos\theta_p \delta\phi'+\lambda\sin\theta_p$. By solving Eqs.\eqref{a5} and \eqref{a6} with the approximation $\lambda \theta^* \cos\theta_p \delta\phi\approx 0$, we can get
\begin{align}
\theta^* = \frac{-H_{S}\cos\theta_p}{4\pi M_s(1+\lambda\sin\theta_p)}.\label{a7}
\end{align}
By using Eq.\eqref{a7} in Eqs.\eqref{a5} and \eqref{a6}, we can get
\begin{align}
\delta\phi' = -\frac{H_{S}\sin\theta_p}{(H_k+4\pi M_s)(1+\lambda\sin\theta_p)}. \label{a8}
\end{align}
Hence,
\begin{align}
	\phi^* = \frac{3\pi}{2} -\frac{H_{S}\sin\theta_p}{(H_k+4\pi M_s)(1+\lambda\sin\theta_p)}.\label{a9}
\end{align}

$\theta^*=\pi/2$, Eq.\eqref{a4} and Eq.\eqref{a7}, Eq.\eqref{a9} represent the steady states of the magnetization when the current is below $I_{min}^c$ and above $I_{max}^c$ respectively.

\subsection{Impact of $\alpha$ and $M_s$ on the frequency}
\begin{figure}
	\centering\includegraphics[angle=0,width=1.0\linewidth]{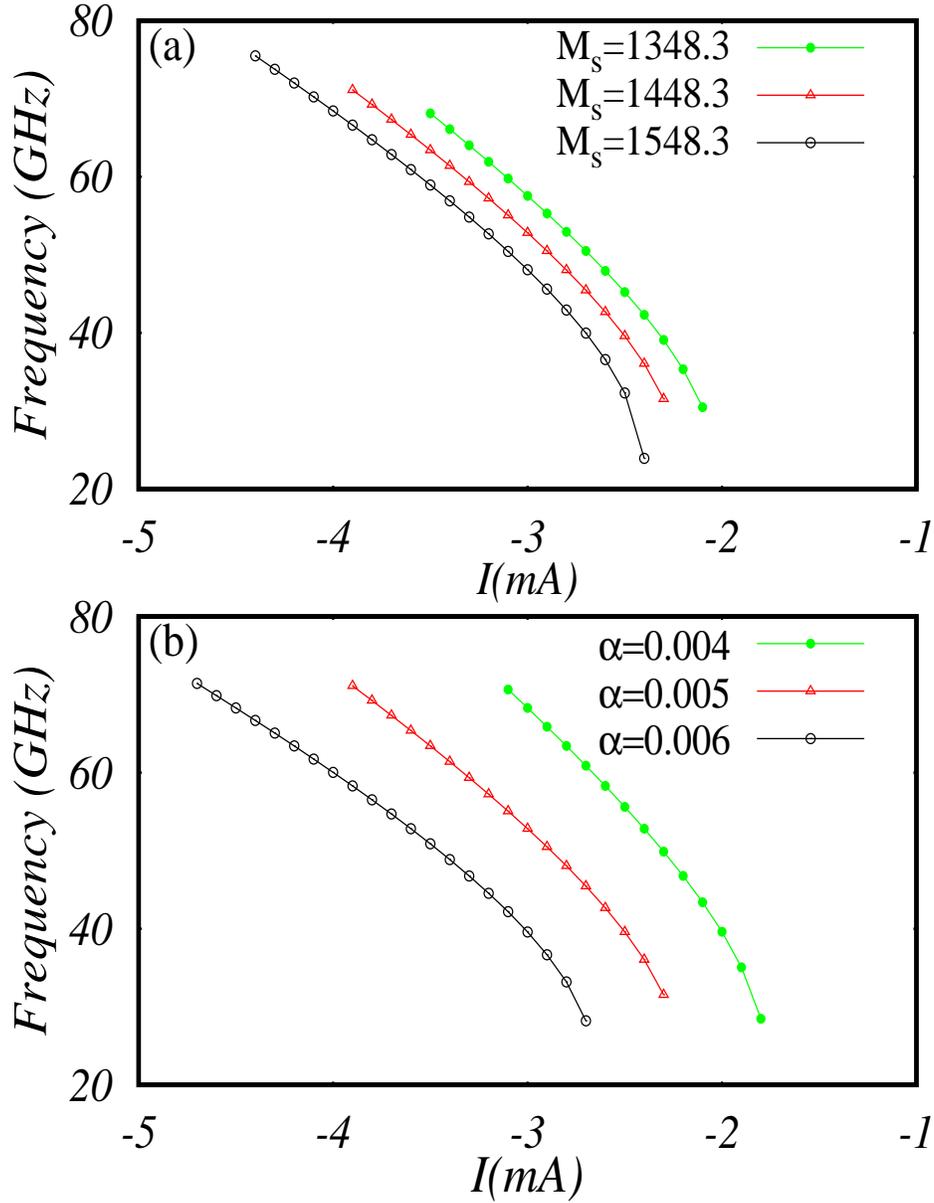}
	\caption{(Color online) Variation of the frequency of STNO's oscillations with respect to current for different values of (a) $M_s$ when $\alpha$ = 0.005  and (b) $\alpha$ when $M_s$ = 1448.3 emu/cc in the absence of thermal noise.}
	\label{fig6}
\end{figure}

To investigate the impact of the material parameters on STNO, the frequency of the oscillations is plotted for different values of saturation magnetization($M_s$) and damping parameter($\alpha$) against current in Fig.\ref{fig6}(a) and (b) respectively.  Since the present work deals with cobalt material for the free layer, Figs.\ref{fig6} are plotted for three different values of $M_S$(1348.3, 1448.3 and 1548.3 emu/cc) and three values of $\alpha$(0.004, 0.005 and 0.006).   From Figs.\ref{fig6},  it can be understood that for fixed values of current the frequency increases with a decrease of $\alpha$ and $M_s$. Fig.\ref{fig6}(a) confirms that high frequency is acheived at large current when the saturation magnetization is large.  On the other hand, Fig.\ref{fig6}(b)   implies that high frequency can be achieved at low currents when the valueof damping parameter is low.  We have also observed corresponding impact with other parameters such as spin polarization efficiency($\eta$) and anisotropy field($H_k$).

\subsection{Impact of thermal noise}
	To investigate the impact of thermal noise on the STNO, the dynamics of magnetization is studied by the effective field with thermal noise as ${\bf H}_{eff}=H_k m_x {\bf e}_x - 4\pi M_s m_{z}{\bf e}_z + {\bf H}_{th}$, where the thermal noise ${\bf H}_{th}$ is given by~\cite{roma,smirnov,Hahn}
	\begin{align}
	{\bf H}_{th} = \sqrt D~ {\bf G} \label{Hth},~~~D =   {\frac{2\alpha k_B T}{\gamma M_s \mu_0 V \triangle t}}.
	\end{align}
	Here, ${\bf G}$ is the Gaussian random number generator vector of the STNO with components $(G_{x}, G_{y}, G_{z})$, which satisfies the statistical properties $<G_{m}(t)>=0$ and $<G_{m}(t) G_{n}(t')>=\delta_{mn}\delta(t-t')$ for all $m,n=x,y,z$.  $k_B$ is the Boltzmann constant, $T$ is the temperature, $\triangle t$ is the step size of the time scale used in the simulation and $\mu_0$ is the magnetic permeability at free space.  

	\begin{figure*}
		\centering\includegraphics[angle=0,width=1\linewidth]{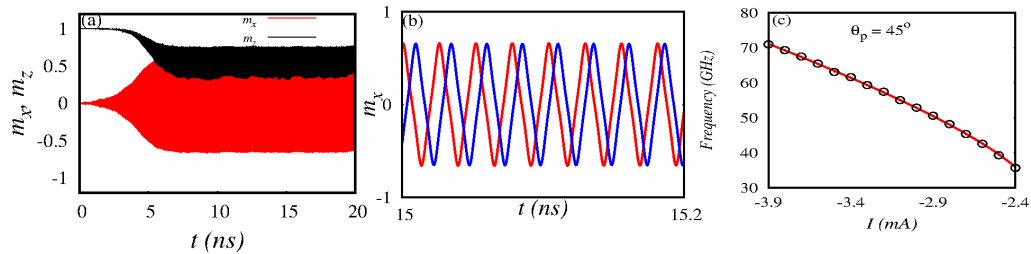}\
		\caption{(Color online) (a) Time evolution of (a) $m_x$ and $m_z$ for $T=300$ K and $I$ = -2.5 mA.  (b) Time evolution of $m_x$ in the presence(red solid line) and absence(blue solid line) of thermal noise. (c) Frequency with respect to current in the absence(red solid line) and presence(black open circle) of thermal noise when $\theta_p = 45^\circ$.}
		\label{fig7}
	\end{figure*}
	Fig.\ref{fig7}(a) shows the evolution of $m_x$ and $m_z$ with respect to time for $T=300$ K and $I$ = -2.5 mA. The irregularities in the time evolution arises due to the thermal noise (see Fig.\ref{fig1}).  Fig.\ref{fig7}(b) shows the time evolution of $m_x$ in the presence (solid red line) and absence (solid blue line) of thermal noise for the same initial conditions, which confirms a small phase shift due to the thermal noise. From Fig.\ref{fig7}(c) it is verified that there is no appreciable variation in the frequency of oscillations of magnetization due to thermal noise.

\end{document}